\documentclass[a4paper,fleqn,usenatbib]{mnras}
\usepackage{amsmath,graphicx,amssymb}
\usepackage[varg]{txfonts}
\usepackage{ae,aecompl}
\newcommand{\vel}{{v}}

\newcommand{\zav}[1]{\left(#1\right)}
\newcommand{\hzav}[1]{\left[#1\right]}

\title{Torsional oscillations and observed rotational period variations in
early-type stars}

\author[J. Krti\v{c}ka et al.]{J.~Krti\v{c}ka,$^{1}$\thanks{E-mail:
krticka@physics.muni.cz (JK)} Z.~Mikul\'a\v sek,$^{1}$
G.\,W.\,Henry$^{2}$, P.~Kurf\"urst$^{1}$ and M.\,Karlick\'y$^{3}$
\\
$^{1}$Department of Theoretical Physics and Astrophysics,
Masaryk University, Kotl\'a\v rsk\' a 2, CZ-611\,37 Brno, Czech
Republic\\ 
$^{2}$Center of Excellence in Information Systems, Tennessee
State University, Nashville, Tennessee, USA\\
$^{3}$Astronomical Institute, Academy of Science of CR, Fri\v{c}ova 298,
        CZ-251 65, Ond\v rejov, Czech Republic
}

\date{Received}

\pubyear{2016}

\begin{document}
\label{firstpage}
\pagerange{\pageref{firstpage}--\pageref{lastpage}}
\maketitle

\begin{abstract}
Some chemically peculiar stars in the upper main sequence show 
rotational period variations of unknown origin. We propose these variations 
are a consequence of the propagation of internal waves in magnetic 
rotating stars that lead to the torsional oscillations of the star.
We simulate the magnetohydrodynamic waves and calculate resonant frequencies 
for two stars that show rotational variations: CU~Vir and HD~37776. We 
provide updated analyses of rotational period variations in these stars and 
compare our results with numerical models. For CU~Vir, the length of the 
observed rotational-period cycle, $\mathit\Pi=67.6(5)$\,yr, can be well 
reproduced by the models, which predict a cycle length of 51\,yr. However, 
for HD~37776, the observed lower limit of the cycle length, 
$\mathit\Pi\geq100$\,yr, is significantly longer than the numerical models 
predict.  We conclude that torsional oscillations provide a 
reasonable explanation at least for the observed period variations in 
CU~Vir.
\end{abstract}


\section{Introduction}

Chemically peculiar (CP) stars in the upper main sequence show light variability
attributed to rotational modulation of magnetic spots of differing surface
abundances.  Radiative flux redistribution in the spots results from various
bound-free \citep[ionization,][]{peter,lanko} and bound-bound atomic
\citep[line,][]{vlci,ministr,molnar} transitions. Surface abundance maps derived
from Doppler imaging \citep[e.g.,][]{leuma,silkow} can be used to predict the
light variability in CP stars \citep[e.g.,][]{myteta,prvalis}.

The brightness variability in CP stars allows measurement of their rotational
periods.  The strict periodicity observed in their light curves allows precise
determination of their periods with typical relative uncertainties of the order
$10^{-6}$\,--\,$10^{-5}$ \citep[e.g.,][]{adelm}. This facilitates the
search for very minute changes in the rotational periods. For single CP stars,
the usual mechanisms of period change are related to stellar evolution.
Unfortunately, evolutionary changes in stellar rotation \citep{rotmod} are not
detectable in main-sequence stars \citep{mikmos}. As a result, most CP stars
have very constant rotational periods.

However, there are exceptions. The hottest CP stars with surface magnetic
fields and winds may show rotational braking as a result of angular momentum 
loss via the magnetized stellar wind \citep{brzdud}. This effect was 
discovered in the helium-rich star $\sigma$~Ori~E by \citet{town}. Period 
variations in HD~37776 discovered by \citet{mik901} were also attributed to 
angular momentum loss. However, subsequent analysis of HD~37776 by 
\citet{zmenper2} revealed a significant cubic term in the star's ephemeris, 
inconsistent with simple rotational braking. In the CP star CU~Vir, intervals 
of rotational braking were found to alternate with intervals of rotational 
acceleration by \citet{zmenper2}. 

These new findings in stars HD~37776 and CU~Vir need to be explained. Here we
study the torsional oscillations that result from the interaction of
rotation and magnetic field \citep[see][for a similar idea]{step}. Although
originally introduced for other purposes (cf., \citealt{mestel}, pages 161--163,
see also \citealt{mw}), we show that the torsional oscillations are able to
explain the period variation in CU~Vir.

\section{Torsional oscillations}
\label{torsoc}

We assume that the star is in equilibrium given by the solution of the
ideal magnetohydrostatic equation \citep[e.g.,][page 314]{biblerot}
\begin{equation}
0=-\nabla p +\frac{1}{4\pi}(\nabla\times\boldsymbol{B})\times\boldsymbol{B}+
\rho\boldsymbol{g}_\text{eff}.
\end{equation}
The equation is written in a reference frame connected with the star rotating
rigidly with angular frequency $\Omega$. Therefore, $\boldsymbol{g}_\text{eff}$
stands for effective gravity that results from the gravitational and centrifugal
accelerations, and $\boldsymbol{B}$ is the magnetic field.

We will study the incompressible waves in such a star. The velocity and magnetic
field perturbation $\delta\boldsymbol{\vel}$ and $\delta\boldsymbol{B}$ shall
fulfill the linearized magnetohydrodynamic equations in the form of
\citep[c.f.,][]{asaci}
\begin{align}
\nabla\cdot\delta\boldsymbol{\vel}&=0,\\
\label{momlin}
\rho\frac{\partial\delta\boldsymbol{\vel}}{\partial t}&=
\frac{1}{4\pi}(\nabla\times\delta\boldsymbol{B})\times\boldsymbol{B}+
\frac{1}{4\pi}(\nabla\times\boldsymbol{B})\times\delta\boldsymbol{B},\\
\label{indlin}
\frac{\partial\delta\boldsymbol{B}}{\partial t}&=
\nabla\times(\delta\boldsymbol{\vel}\times\boldsymbol{B}).
\end{align}
Taking partial derivative of Eq.~\eqref{momlin} with respect to time and
using Eq.~\eqref{indlin} we derive the wave equation
\begin{equation}
\label{supervlna}
\rho\frac{\partial^2\delta\boldsymbol{\vel}}{\partial t^2}=
\frac{1}{4\pi}\{\nabla\times[\nabla\times
(\delta\boldsymbol{\vel}\times\boldsymbol{B})]\}\times\boldsymbol{B}+
\frac{1}{4\pi}(\nabla\times\boldsymbol{B})\times
[\nabla\times(\delta\boldsymbol{\vel}\times\boldsymbol{B})].
\end{equation}

As an application of Eq.~\eqref{supervlna}, one can assume an axisymmetric system 
following \citet[pages 161--163, see also \citealt{mw}]{mestel} with
$\delta\boldsymbol{\vel}\equiv(\vel_R,\vel_\varphi,\vel_z)=(0,R\delta\Omega,0)$
and $\boldsymbol{B}\equiv(B_R,B_\varphi,B_z)=(B_R,0,B_z)$
in cylindrical coordinates $R$, $\varphi$, and $z$.
With constraint $\nabla\cdot\boldsymbol{B}=0$ this leads to the wave
equation in the form of \citep{mw}
\begin{equation}
\label{vlnrovmes}
\rho\,R^2\frac{\partial^2\delta\mathit{\Omega}}{\partial t^2}=\frac{1}{4\pi}\,
\boldsymbol{B}\cdot\nabla\hzav{R^2
\zav{\boldsymbol{B}\cdot\nabla}\delta\mathit{\Omega}}.
\end{equation}
We note that the second term on the right-hand side of Eq.~\eqref{supervlna} is
zero because it is given by the vector product of two parallel vectors.

Simple magnetic field configurations are unstable inside the stars
\citep[e.g.,][]{bra07} and stable internal field is composed of poloidal and
toroidal field \citep{brano}. However, there is no analytical description of
such complex fields. Consequently, to proceed in the analysis of torsional
oscillations in magnetic stars, we selected simple magnetic field
configurations, which may not fully describe the internal field.
%The simplest
%configuration of uniform magnetic field oriented parallel to the rotational axis
%$\boldsymbol{B}\equiv(B_R,B_\varphi,B_z)=(0,0,B)=\text{const.}$ does not
%provide a resonable model for torsional oscillations. Alfv\'en waves propagate
%along the rotational axis in such field, and consequently the largest amplitude
%of angular velocity variations appears at the stellar pole ($z=R_*$). Such
%oscillations can be hardly observed. Similar problem appears for axysimmetric
%configurations in the regions close to the rotational axis.

The general wave equation describing the torsional oscillations
Eq.~\eqref{supervlna} is very complicated, because it involves double curl
operator and double vector product. The resulting solution describes 3D
oscillations of a star. Therefore, the solution of this equation is cumbersome
even for numerical models. However, only the azimuthal component of
Eq.~\eqref{supervlna} is relevant to study the rotational period variations in
magnetic stars. Consequently, the easiest way how to proceed is to find such
magnetic field configuration, for which only the azimuthal component of the
right hand side of Eq.~\eqref{supervlna} is nonzero. Due to the first term on
the right hand side of Eq.~\eqref{supervlna} including the vector product this
can be achieved for magnetic fields with $B_\varphi=0$ (in cylindrical
coordinates $R$, $\varphi$, and $z$). The simplest form of wave equation
contains the derivatives with respect to only one spatial variable (e.g., $R$),
for which the angular velocity perturbations are functions of only one spatial
variable and time. This is fulfilled for magnetic fields for which the radial
component is a function of $R$ only. The constraint
$\nabla\cdot\boldsymbol{B}=0$ then yields the magnetic field in the form of
$\boldsymbol{B}\equiv(B_R,B_\varphi,B_z) =(B(R),\,0,\,
-{z}/{R}\,{\text{d}\!\zav{RB(R)}}/{\text{d}R})$. 

We assume magnetic field that has the above mentioned form,
\begin{equation}
\label{magpole}
\boldsymbol{B}\equiv(B_R,B_\varphi,B_z)=\zav{B,0,-\frac{zB}{R}},
\end{equation}
where $B$ is constant\footnote{%
The magnetic field is oriented outwards (or inwards) in both stellar poles. This can be avoided
assuming the narrow current sheet in the region $z\in\left<-L/2,L/2\right>$ with
$L\ll R_*$. The magnetic field is then
$\boldsymbol{B}\equiv(B_R,B_\varphi,B_z)=(B,0,-zB/R)$ for $z>L/2$ and
$\boldsymbol{B}\equiv(B_R,B_\varphi,B_z)=(-B,0,zB/R)$ for $z<-L/2$. 
This does not change the form of the resulting wave equation for $L
\rightarrow 0$.
}.
Assuming the angular velocity perturbations in the form of
$\delta\boldsymbol{\vel}\equiv(\vel_R,\vel_\varphi,\vel_z)=(0,R\delta\Omega,0)$
the only nonzero component of Eq.~\eqref{supervlna} gives the wave equation in
the form of
\begin{equation}
\label{vlnrov}
\rho\,R^2\frac{\partial^2\delta\mathit{\Omega}}{\partial t^2}=\frac{B^2}{4\pi}\,
\frac{\partial}{\partial R}\zav{R^2\frac{\partial\delta\mathit{\Omega}}
{\partial R}}.
\end{equation}
For $R\,{\partial^2\delta\mathit{\Omega}}/{\partial R^2}\gg
{\partial\delta\mathit{\Omega}}/{\partial R}$ the wave speed is equal to the
Alfv\'en speed, $\varv_\text{A}=B/\!\sqrt{4\pi\rho}$. The field lines of
the adopted magnetic field Eq.~\eqref{magpole} lie in the equatorial plane for
$z=0$ and the more they lie above (or below) the equatorial plane the more they
diverge from the horizontal plane $z=\text{const.}$ The magnetic field has
therefore a goblet-like structure. The Alfv\'en waves propagate along the field
lines and cause the torsional oscillations that are perpendicular to the
magnetic field. The field lines of Eq.~\eqref{magpole} are also normal to the
stellar surface at the equatorial plane, therefore the amplitude of torsional
oscillation is large in this region. The adopted magnetic field has an internal
stellar field that is similar in strength to the surface field, which is not the
case of dipole field model.

The wave equation \eqref{vlnrov} preserves its form for the scale
transformations $\boldsymbol{B}'=\gamma\boldsymbol{B}$ and
$t'=\gamma^{-1}t$, where $\gamma$ is a constant. This means that the solution
of the wave equation for a given star is the same as the solution for a
magnetic field scaled by factor $\gamma$ at a time scaled by $\gamma^{-1}t$.

A similar problem for more complex fields also leads to the wave equation and
resonant modes described in the following text. Wave equations similar to
Eq.~\eqref{vlnrov} can be obtained also for homogeneous magnetic field
$\boldsymbol{B}=\text{const.}$ or for magnetic field of \citet[see
\citealt{bra07}]{rob}, that in spherical coordinates reads $\boldsymbol{B}\equiv
(B_r,B_\vartheta,B_\varphi)= (B/2\cos\theta(5-3r^2/R_*^2),
B/2\sin\theta(6r^2/R_*^2-5),0)$ with $B$ being constant.

\section{Numerical simulation of torsional oscillations}

For numerical simulations we have rewritten Eq.~\eqref{vlnrov} in a
flux-conservative form
\begin{equation}
\label{jedvln}
\frac{\partial \boldsymbol{u}}{\partial t}=\mathsf{F}\,\frac{\partial
\boldsymbol{u}}{\partial R},
\end{equation}
where $\boldsymbol{u}\equiv(u_1,u_2)^\text{T}=
({\partial\delta \mathit\Omega}/{\partial t},\,
R^2B\,{\partial\delta \mathit\Omega}/{\partial R})^\text{T}$ and
$\mathsf{F}$ is a matrix
\begin{equation}
\mathsf{F}=\zav{
\begin{array}{cc}
0 & \displaystyle\frac{B}{4\pi\rho R^2}\\
R^2B & 0\\
\end{array}}.
\end{equation}
The density distribution for the solution of Eq.~\eqref{jedvln} was taken from
the MESA stellar evolutionary models \citep{mesa1,mesa2}. The boundary
conditions were selected in a same fashion as boundary conditions for
stellar pulsations \citep{biblerot}. We anticipate that $\delta\mathit\Omega$ is
constant close to the stellar surface, therefore we assume a solid wall boundary
condition for $\partial\delta \mathit\Omega/{\partial R}$ (or $u_2$), whereas we
assume reflecting boundary condition for the inner boundary condition. Because
the variables $u_1$ and $u_2$ are connected with a derivative, we selected the
boundary conditions for $u_1$ in the opposite manner, i.e., reflecting boundary
condition at the outer boundary and solid wall boundary condition at the inner
boundary. We introduce the external forcing at the inner boundary, which seems
to be more physically meaningful.

The model equations are solved for $z=0$. The wave equation is independent
of $z$, however the boundary conditions and consequently also the derived
resonance frequencies depend on $z$. We discuss the implications of this
assumption in Sect.~\ref{disc}.

We solved Eq.~\eqref{jedvln} numerically using the leapfrog method 
\citep{nrc}. We used $N$ equidistantly spaced grid points $R_1,\, R_2,\dots,
R_N$ within the stellar 
radius. We assumed no initial perturbation, $\boldsymbol{u}(R,t=0)=0$. The 
inner boundary conditions (at $R=R_1$) with external forcing were represented as
\begin{align}
u_1(R_1,t)&=-u_1(R_2,t)+A_1\sin(\omega t),\\
u_2(R_1,t)&=u_2(R_2,t)-A_2\sin(\omega t),
\end{align}
where $\omega$ is the frequency of external forcing and $A_1$ and $A_2$ its
amplitudes. The subscripts of $u$ denote its components, and the subscript
of the radius denotes individual grid points. We selected $R_1$ corresponding 
to the inner convection zone as derived from the MESA models. The outer 
boundary conditions (at $R=R_N$) are represented as
\begin{align}
\label{u1okraj}
u_1(R_{N},t)&=u_1(R_{N-1},t),\\
u_2(R_{N},t)&=-u_2(R_{N-1},t).
\end{align}
To avoid instabilities, the Courant-Friedrichs-Lewy condition was used to 
limit the length of the time step. The selected time step was a fraction
of the minimum travel time of Alfv\'en waves across one grid zone.

We used $N=1000$ in our calculations. Because the Alfv\'en waves are very 
fast close to the stellar surface, the time step required by the 
Courant-Friedrichs-Lewy condition is very short, which results in a 
prohibitively short time step for the calculations. To make the calculations 
more tractable, we omit the surface layers from our calculations. Since this 
typically corresponds to a very small fraction of the stellar radius, this 
modification does not strongly affect the final results. Because the wave 
equation \eqref{vlnrov} is linear, the absolute scale of the external 
forcing amplitudes $A_1$ and $A_2$ is unimportant for the subsequent 
numerical analysis.

\section{Application of numerical simulations to individual stars}

\subsection{CU Vir}
\label{cuvirap}

An abrupt period increase in CU~Vir (HD 124224) was reported by
\citet{pyper1,pyper2}. On the other hand, \citet{zmenper2} found evidence in the
available photometric and spectrophotometric measurements for gradual long-term
period variations \citep[see also][]{step} and even detected intervals where the
period was decreasing. We assumed the latter model of the period variations.
For our analysis we selected the MESA solar-metallicity evolutionary model with
an initial mass of $3\,M_\odot$ at the age of $9\times10^7\,\text{yr}$, which
corresponds to CU~Vir parameters derived by \citet{kobacu}. We adopted
$B_\text{p}=1\,\text{kG}$, which is representative of CU~Vir's surface
\citep{kocuvir}. The radial boundaries of the computational zone are
at $R_1=0.159\,R_*$ and $R_{N}=0.997\,R_*$,
where $R_*$ is the stellar radius as calculated by MESA models.

\begin{figure}
\centering
\resizebox{\hsize}{!}{\includegraphics{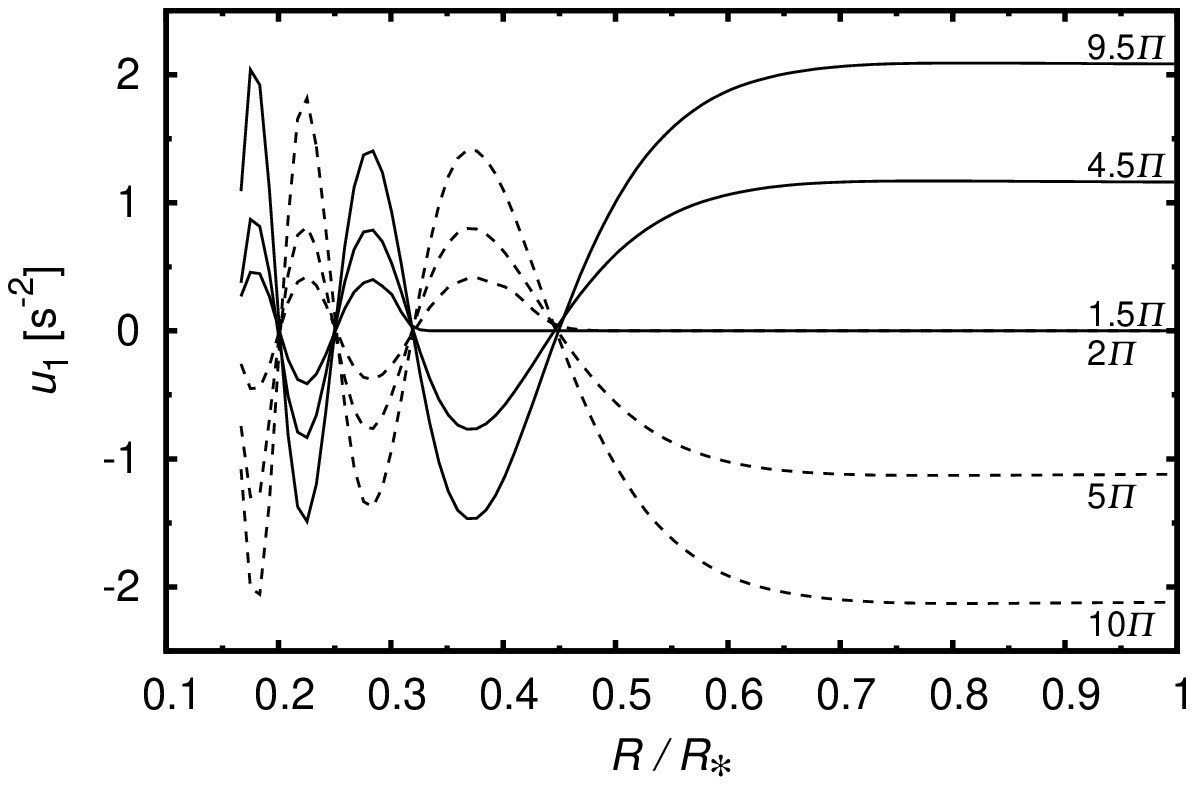}}
\resizebox{\hsize}{!}{\includegraphics{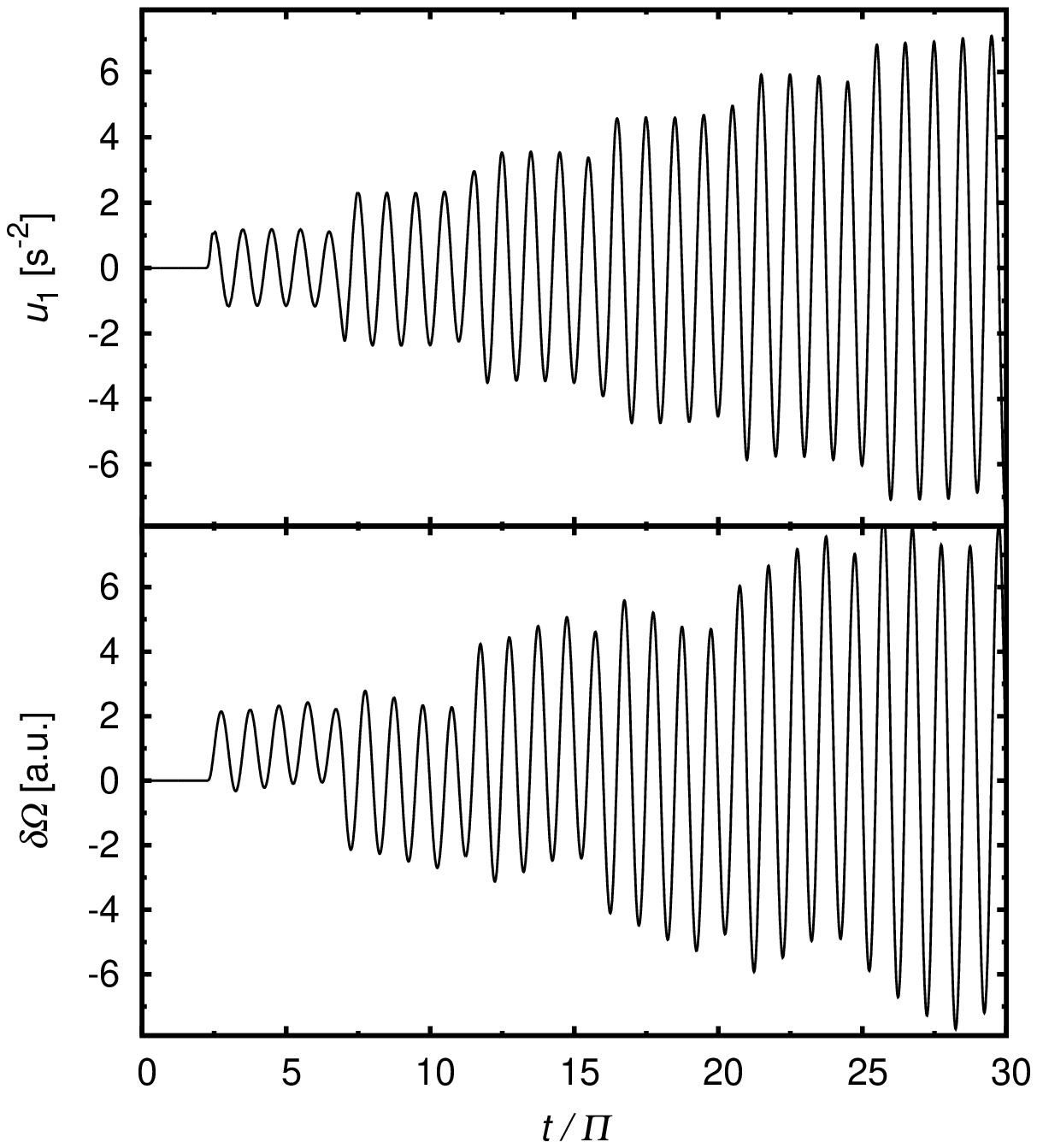}}
\caption{The waves connected with torsional oscillations in the model stellar
interior of CU~Vir
for $\omega=3.76\times10^{-8}\,\text{s}^{-1}$, which corresponds to resonance. {\em Upper panel}: The variable 
$u_1$ at different stages of the wave evolution. The curves are denoted by the 
elapsed time in units of the period of external forcing 
$\mathit\Pi=2\pi/\omega\approx5.3\,\text{yr}$. 
{\em Middle panel}: The behavior of $u_1$ at the
outer boundary of the model. {\em Lower panel}: The evolution
of $\delta\Omega$ derived from the numerical integration of $u_1$.}
\label{cuvirp5}
\end{figure}

In Fig.~\ref{cuvirp5} we plot the solution to the wave equation \eqref{vlnrov}
for $\omega=3.76\times10^{-8}\,\text{s}^{-1}$ (corresponding to a resonate
overtone, see below), $A_1=1\,\text{s}^{-2}$, and
$A_2=1\,\text{cm}^{1/2}\,\text{g}^{1/2}\,\text{s}^{-2}$. The upper panel shows
wave solutions that grow in amplitude. The wave form remains roughly constant
with four knots for a specific value of $\omega$. The lower panel shows the
outer boundary oscillating with a frequency $\omega$ and a growing amplitude.

The observed frequencies are those with the largest growth rates and are 
eigenfrequencies of the corresponding oscillator. Therefore, we scanned all 
frequencies in the range of 
$\omega=10^{-9}\,\text{s}^{-1}$\,--\,$10^{-7}\,\text{s}^{-1}$ 
to find the maximum growth rate of the oscillations. The selected frequencies 
roughly correspond to those detectable from observations with period 
$\mathit\Pi=2$\,--\,$200\,\text{years}$. We allowed the system to evolve for 
$T=600\,$years. The magnitude of the oscillations is estimated from the 
variable $\langle u_1^2\rangle$, defined as a mean value of the square of 
$u_1$ averaged over the period $\mathit\Pi=2\pi/\omega$,
\begin{equation}
\label{ampl}
\langle u_1^2\rangle=\frac{1}{\mathit\Pi}\int_{T-\mathit\Pi}^{T}u_1^2(R_{N},t)\,\text{d} t.
\end{equation}

\begin{figure}
\centering
\resizebox{\hsize}{!}{\includegraphics{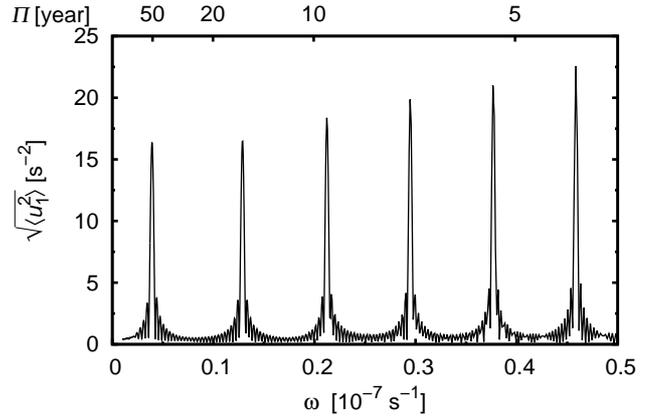}}
\caption{The dependence of the scale mean-squared wave amplitude at the 
stellar surface (see Eq.~\eqref{ampl}) on the frequency after $T=600\,$years 
of wave evolution for CU~Vir. The upper axis is labeled with corresponding 
oscillation periods.}
\label{cuvirven}
\end{figure}

In Fig.~\ref{cuvirven} we plot the results of the simulations. The amplitude of
$\langle u_1^2\rangle$ has several sharp peaks at the basic frequency of the
oscillator and its overtones. There is a maximum amplitude at the period of
about 51~years with overtones at higher frequencies. The first overtones
correspond to periods 15, 9, 7, and 5 years. The basic period of 51~yr is
roughly four times as long as the Alfv\'en wave travel time between the
boundaries of the computational zone, which is 15~years. Therefore, in the basic
mode the computational zone accommodates one fourth of the wave with a node at
the inner boundary and an antinode at the outer boundary. We note that it might
be better to scale $\langle u_1^2\rangle$ by $\omega^{-1}$ in
Fig.~\ref{cuvirven} to account for an $\omega$ dependence of power on external
forcing Eq.~\eqref{u1okraj}, but this has little impact on the final results,
because the derived peaks are relatively sharp.

\subsection{HD 37776}

Rotational period variations in HD~37776 (V901~Ori) were detected 
by \cite{mik901}. A subsequent study by \citet{zmenper2} provided additional 
support for complex period variations in this star. For our analysis, we 
selected the MESA solar-metallicity evolutionary model with an initial mass 
of $8\,M_\odot$ and an age of $3\times10^6\,\text{yr}$. This corresponds
to HD~37776 parameters \citep{otevrenky,mik901}. The star has a very complex 
surface magnetic field \citep{hd3776mag}; consequently we selected 
$B_\text{p}=10\,\text{kG}$, which roughly corresponds to the observed values.  
We selected $R_1=0.203\,R_*$ and $R_{N}=0.983\,R_*$ for the outer model radius,
where $R_*$ is the stellar radius as calculated by MESA models.  

\begin{figure}
\centering
\resizebox{\hsize}{!}{\includegraphics{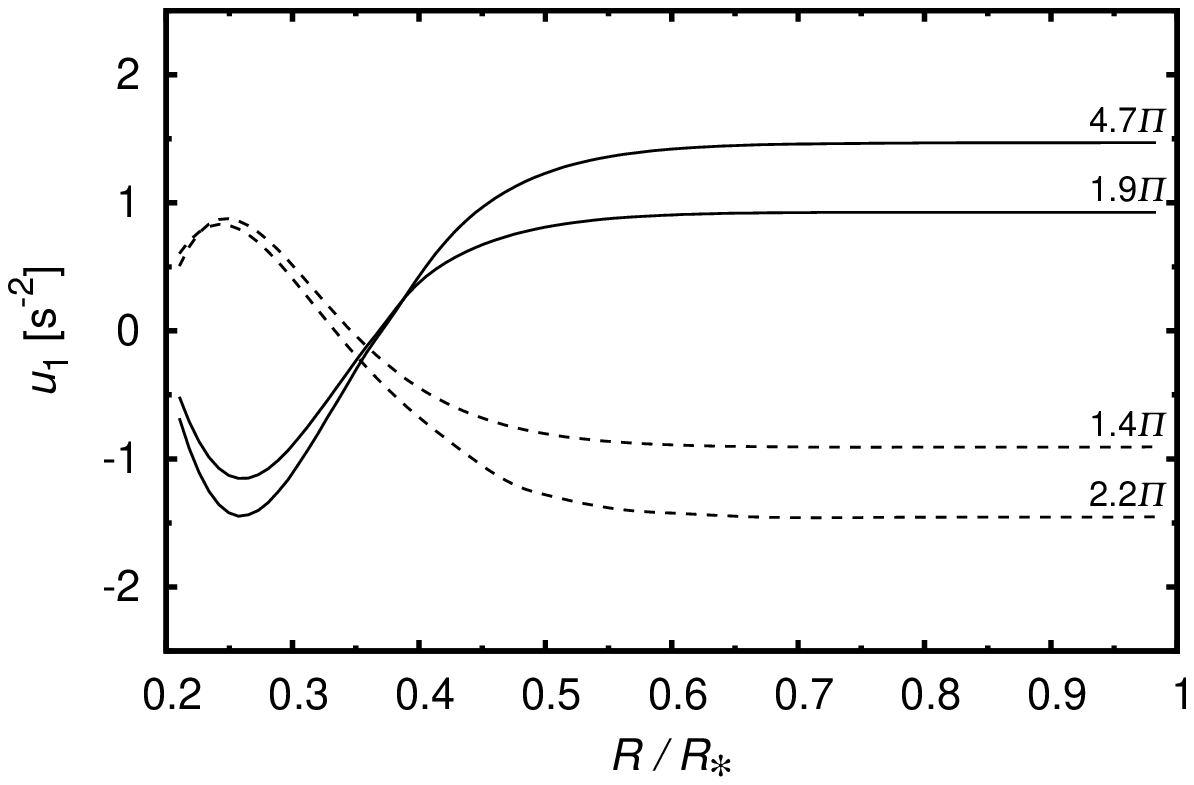}}
\resizebox{\hsize}{!}{\includegraphics{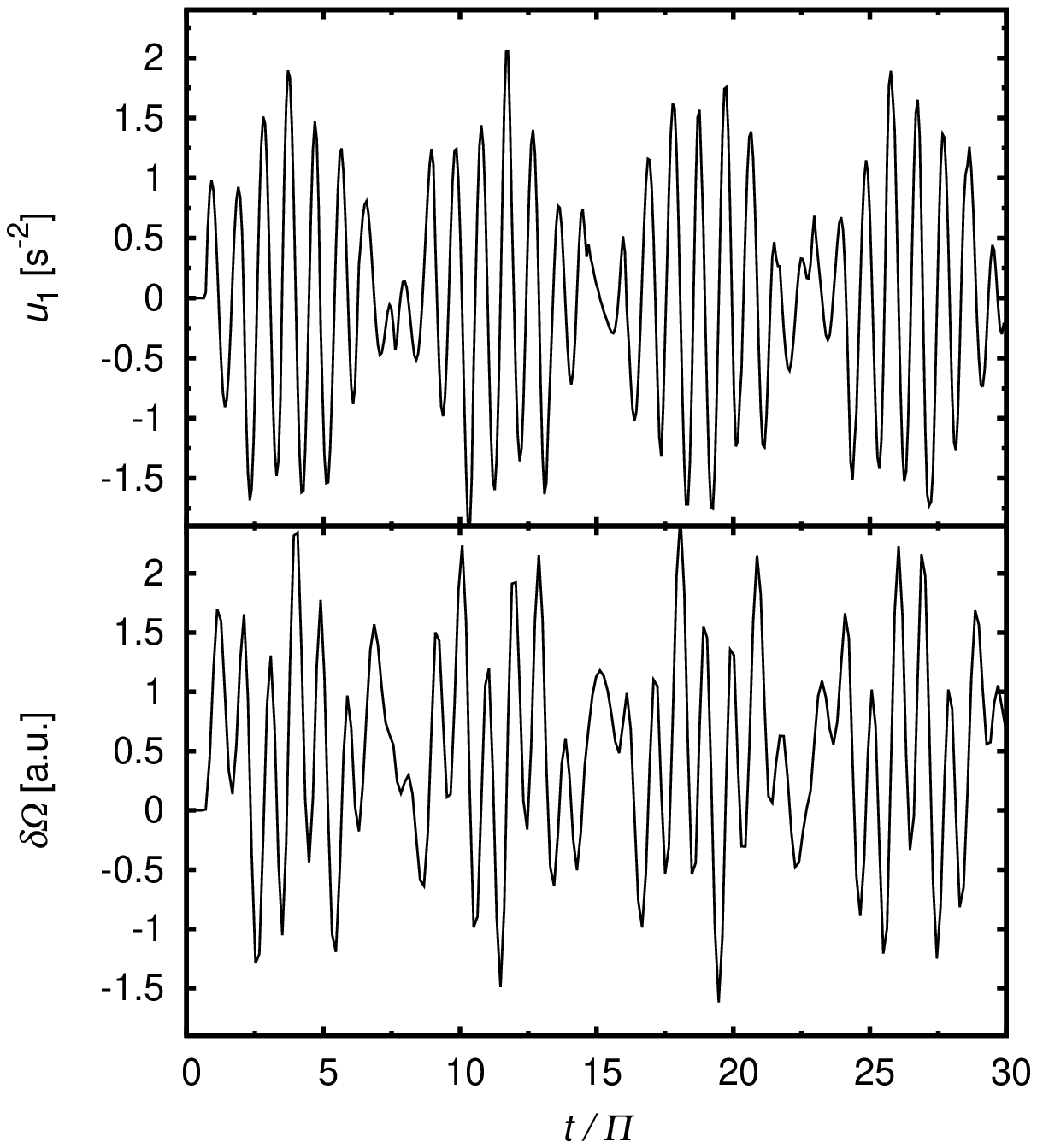}}
\caption{Model of waves connected with torsional oscillations in a stellar
interior corresponding to
HD~37776 for $\omega=1.1\times10^{-7}\,\text{s}^{-1}$, which is slightly out of
the resonance. {\em Upper panel}: The 
variable $u_1$ at different stages of the wave evolution.  The curves are 
denoted by the elapsed time in units of the period of external forcing 
$\mathit\Pi=2\pi/\omega\approx1.8\,\text{yr}$. 
{\em Middle panel}: The behavior of 
$u_1$ at the outer boundary of the model. {\em Lower panel}: The evolution
of $\delta\Omega$ derived from the numerical integration of $u_1$.}
\label{hd37776o7}
\end{figure}

In Fig.~\ref{hd37776o7} we plot the solution of wave equation \eqref{vlnrov} for
$\omega=1.1\times10^{-7}\,\text{s}^{-1}$, $A_1=1\,\text{s}^{-2}$, and
$A_2=1\,\text{cm}^{1/2}\,\text{g}^{1/2}\,\text{s}^{-2}$. The selected frequency
is slightly out of the resonance. The upper panel shows a wave solution with
variable amplitude. The wave form remains roughly constant with one knot (for a
given value of $\omega$). The lower panel of Fig.~\ref{hd37776o7} shows that the
outer boundary oscillates with frequency $\omega$ and an amplitude affected by
beating between the period of external forcing and the eigenfrequency of the
oscillator.

\begin{figure}
\centering
\resizebox{\hsize}{!}{\includegraphics{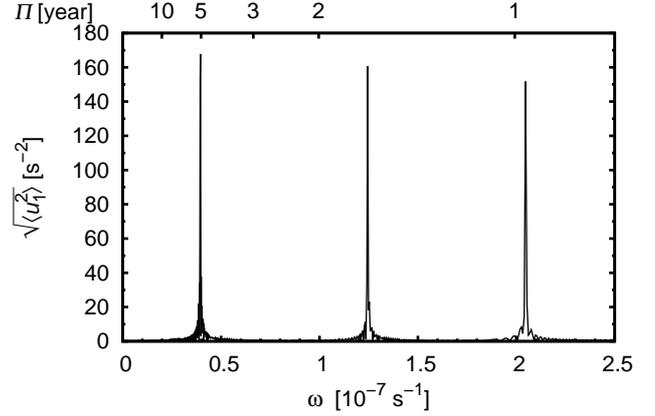}}
\caption{Same as Fig.~\ref{cuvirven}, but for HD 37776.}
\label{hd37776ven}
\end{figure}

We scanned all frequencies in the range
$\omega=10^{-9}\,\text{s}^{-1}$\,--\,$10^{-6}\,\text{s}^{-1}$ to find the
maximum growth rate of the oscillations. We let the system evolve for
$T=600\,$years.  The resulting mean value of the outer boundary amplitude
averaged over the last period of simulations $\langle u_1^2\rangle$ (see
Eq.~\eqref{ampl}) is given in Fig.~\ref{hd37776ven}.  The first three resonance
frequencies correspond to periods of 5, 1.6, and 1 yr. In this case, the basic
5~yr period of torsional oscillations is roughly four times the Alfv\'en wave
travel time between the boundaries of the computational zone, which is
1.6~years. The periods of oscillations are significantly shorter than in
the case of CU~Vir due to stronger surface magnetic field in HD~37776. This
can be seen already from the scaling relations given in Sect.~\ref{torsoc} and
from the results obtained for CU Vir in Sect.~\ref{cuvirap}.

\section{Comparison of numerical results with observations}

To date, rotational period variations of chemically peculiar stars were
phenomenologically described by simple polynomials and their combinations. Our
proposed explanation of the observed period variations requires a new analysis
of the data assuming sinusoidal period variations \citep{mikar11}.

\begin{figure}
\centering
\resizebox{\hsize}{!}{\includegraphics{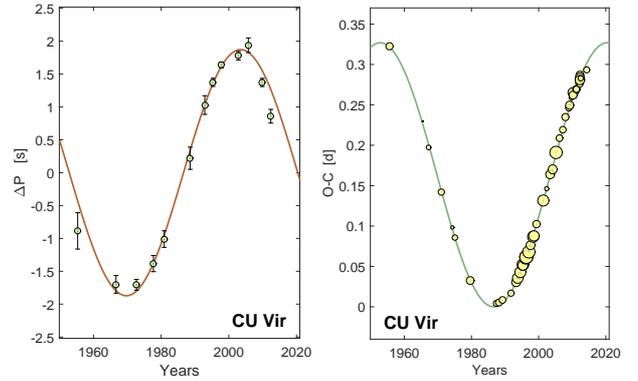}}
\caption{Variations of the rotational period in seconds and phase shifts in 
days for CU\,Vir in the time interval 1949 -- 2015. The observed values are 
fitted by simple sine/cosine model with a period of 
$\mathit{\Pi}=67.6$\,years.}
\label{OCduoCU}
\end{figure}

The variations of the rotational period of CU~Vir are derived from the analysis
of 18\,267 individual photometric, spectroscopic, and radiometric observations
obtained from 1949 to 2015 \citep{zmenper2,mikmos}, supplemented by additional
new $BV$ observations from the T3 0.4\,m Automatic Photoelectric Telescope (APT)
at Fairborn Observatory in southern Arizona. The available observations
together with updated analysis of the period changes will be described in a
separate paper in detail (Mikul\'a\v sek et al., in preparation). We found that
the instantaneous rotational period $P(t)$ varied nearly harmonically according
to the simple relation 
\begin{equation} P(t)=P_0+A\,\sin\zav{2\,\pi\,\frac{t-\mathit{\Theta_0}}{\mathit{\Pi}}},
\end{equation}
where $t$ is a JDhel time, $P_0=0\fd52069424(4)$ is the mean rotational 
period, $A=2\fd167(8)\times10^{-5}=1.872(7)$\,s is the semiamplitude of 
period variations, $\mathit{\Theta_0}=2\,4466\,530(16)=1986.69(5)$, and 
$\mathit{\Pi}=24\,680(190)$\,d $=67.6(5)$\,years, is the length of the period 
cycle $P(t)$. The semiamplitude of the cyclic (O-C) changes 
\begin{equation}
\label{aoc}
A_{\mathrm{OC}}=\frac{A\,\mathit{\Pi}}{2\,\pi\,\overline{P}}
\end{equation}
is $A_{\mathrm{OC}}=0\fd1635=0.31\,\overline{P}$. The period changes and the 
phase shifts in days are presented in Fig.\,\ref{OCduoCU}. Our observations 
extend nearly over the entire cycle of period changes.

The predicted period of torsional oscillations in CU~Vir agrees well with time
scale of period variations observed in this star. Moreover, the period
variations on the scale of ten years \citep{zmenper2} can be nicely explained by
the higher overtones of the torsional oscillations. Consequently, the
torsional oscillations provide a viable explanation of the period variations
found in CU~Vir.

Our second hot CP star with pronounced period variations is 
HD\,37776 (V901\,Ori). Its period analysis is based on 3483 photometric and 
spectroscopic observations obtained between 1976 and 2014 
\citep{zmenper2,mikmos} supplemented by additional new $BV$ observations from 
the T3 0.4\,m APT. We find that 
changes of its instantaneous rotational period of $P(t)$ can be well 
approximated by a parabola or a segment of a harmonic function of period 
$\mathit{\Pi}$ with a maximum at the time $T_{\mathrm{max}}$:
\begin{equation}
P(t)=P_{\mathrm{max}}+A\,\hzav{\cos\zav{2\,\pi\,\frac{t-T_{\mathrm{max}}}{\mathit{\Pi}}}-1},
%\ \quad A_{\mathrm{OC}}=\frac{A\,\mathit{\Pi}}{2\,\pi\,\overline{P}},
\end{equation}
where $P_{\mathrm{max}}$ is the maximum rotational period,
and the semiamplitude of the corresponding cyclic variations of (O-C) is
given by Eq.~\eqref{aoc}.

We are not able to determine the true value of the period cycle $\mathit{\Pi}$,
only its lower limit: $\mathit{\Pi}\geq 100$\,years. Therefore, 
$P_{\mathrm{max}}\cong1\fd538784$, $A\geq1\fd42\times 10^{-4}=12.2$\,s, 
$A_{\mathrm{OC}}\geq0\fd54=0.35\,\overline{P}$, and 
$T_{\mathrm{max}}\geq2006.5$, and HJD$(T_{\mathrm{max}})=2\,453\,910$.  
 
\begin{figure}
\centering
\resizebox{\hsize}{!}{\includegraphics{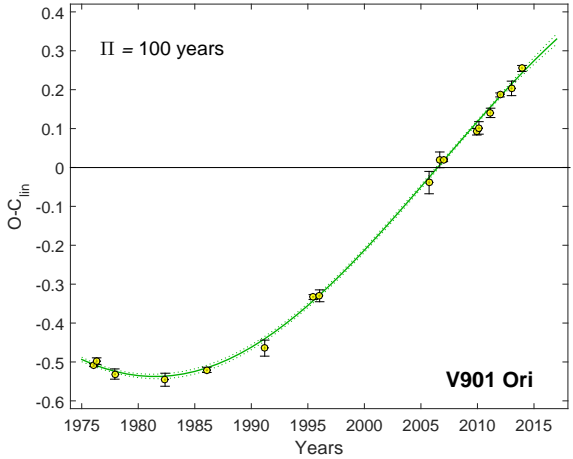}}
\caption{The observed phase shifts in days for V901 Ori fitted with a
sine function with the minimal acceptable period $\mathit{\Pi}=100$\,years.}
\label{OC901}
\end{figure}

For HD~37776 the predicted torsional oscillation period cycle is significantly
shorter than the observed period cycle. This might be explained by a difference
between the surface and inner magnetic field. The surface field of HD~37776 is
very complex \citep{hd3776mag}, and it may decay over time as seen in magnetic
stars found in open clusters \citep{otevrenky}. It is possible that part of the
observed decay may be connected with equalization of the surface and inner
magnetic field. However, it does not follow from the simulations that the
surface field is stronger than the inner field \citep{brano}.

On the other hand, there are still other possibilities that can explain the
observed period variations. Besides that excluded by \citet{mik901}, the most
relevant could be the tidal interaction with an orbiting low-mass body
\citep{ruzovamary}. For example, a possible scenario could include a low-mass
object on a non-synchronized precessing orbit. We postpone this possibility 
for a future detailed study.

\section{Discussion}
\label{disc}

The model presented in this paper includes only the basic physics describing the
problem. Moreover, we solve the wave equation for $z=0$ only. In reality, the
internal magnetic field has a more complex structure, and stable internal
magnetic fields are given by the combination of poloidal and toroidal magnetic
fields (see \citealt{brano} and \citealt{bras}, for a review). Because the wave
equation \eqref{vlnrov} is specified on the field lines a more realistic field
topology would lead to a lengthening of the oscillation period on each field
line \citep[see also][]{link}. As the basic frequency differs on each field
line, the basic frequency of the global mode will be given by a weighted average
of individual frequency modes. The resulting phase mixing \citep[e.g.,][]{hepr}
may lead to the damping of oscillations, but since the individual peaks in the
periodogram are rather broad (see~Fig.~\ref{cuvirven}), we expect that the
oscillations may still be observable.

The magnetic field dominates the atmospheres of studied stars. Therefore, their
surface may resemble the crust of neutron stars forcing the individual field
lines to oscillate with the same frequency. Consequently, the variations of
rotational period of CP stars may have the same explanation as quasi-periodic
oscillations of neutron stars. In this case the frequencies of rotational period
variations may be associated with turning points or edges of the MHD continuum
\citep{levin} or with normal oscillation modes \citep{lee}.

Another problem is connected with existence of two shallow convection
zones associated with local opacity enhancements inside massive stars
\citep[e.g.,][]{hmkon}. However, because these zones contain only a very small
fraction of the stellar mass, we expect that their existence does not
significantly alter our results.

Our model assumes small perturbations of otherwise uniform stellar angular
velocity. This does not seem to be in contradiction with vertical differential
rotation profiles derived from asteroseismology in BA stars
\citep{briketka,lakedistrict}. However, angular momentum transport by, e.g.,
internal gravity waves \citep{rogers} may lead to damping of torsional
oscillations.

It is not clear what mechanism drives the torsional oscillations. The forcing
mechanism may be connected either with the core convection zone or with winds
blowing from the stellar surface. If some Ap stars originate in binary star
mergers \citep{maupaunet,botu}, then subsequent relaxation processes may provide
a convenient mechanism that drives the torsional oscillations. This would mean
that CU~Vir is a merger product. This could possibly provide a consistent
explanation of circumstellar matter that is required to power the pulsed radio
emission observed in CU Vir \citep{trigilio,kellett}. An alternative explanation
of line-driven wind mass-loss rate of about
$10^{-12}\,{M}_\odot\,\text{yr}^{-1}$ \citep{letcuvir} is problematic due to
the low effective temperature of the star \citep{metuje}.

The wave equation \eqref{vlnrov} is linear and so allows oscillations with 
arbitrary amplitudes. This is unrealistic in nature, and several processes 
may lead to damping of strong oscillations. One obvious process is 
reconnection, which may occur in the case of oscillations that are wound up 
too much. The oscillations may also be damped due to vertical shear.

Another important question is connected with the fact that period variations are
detected in only a handful of magnetic stars, while the light curves of most of
them are adequately described with constant periods
\citep[e.g.,][]{adelm,jednod,wr,bernard}. While a lack of suitable long
observing series may provide an obvious answer for many stars, more subtle
reasons may be missing. However, since a similar question is not clearly
resolved in many notorious classical pulsating stars, i.e., the question why
there are stars that do not pulsate in the instability strip
\citep[e.g.,][]{fontana,balpul,saio}, we postpone this question for a future
study.

\section{Conclusions}

We simulated stellar torsional oscillations that result from the interaction of
the internal magnetic field and differential rotation. The simulations were
calculated for the chemically peculiar stars CU~Vir and HD~37776, which both
have rotational period variations. We derived the internal structure of
individual modes and calculated the wave resonance frequencies, for which the
amplitudes of surface angular frequency variations are the largest. For each
star we found a basic frequency and several high-order overtones.

We provide a new analysis of period variations in the stars CU~Vir and HD~37776
assuming periodic rotational period variations. For CU~Vir, the length of the
rotational period cycle $\mathit\Pi=67.6(5)$\,yr can be well reproduced by
numerical models, which predict a cycle length of 51\,yr. The numerical model
also predicts the variations on the scale of about 10~yr in agreement with
observations. Consequently, the torsional oscillations provide a reasonable
explanation of the observed period variations of CU~Vir. On the other hand, for
HD~37776 the observed lower limit of the period cycle, $\mathit\Pi\geq100$\,yr,
is significantly longer than the predicted cycle length of 5\,yr. It is
immediately clear from the scaling of the wave equation with the magnetic field
and from the observed strength of the field that the model cannot reproduce the
observations of both stars. There may be other possible explanations for the
observed period variations in HD~37776.

\section*{Acknowledgements}
This work was supported by grants GA \v{C}R 16-01116S and P209/12/0103.
GWH acknowledges support from Tennessee State University and the State of
Tennessee through its Centers of Excellence program.

\end{document}